\documentclass[showpacs,amsmath,amssymb,aps,showkeys,floatfix,prd,a4paper]{revtex4}

\usepackage[dvips]{graphicx}
\usepackage{dcolumn}
\usepackage{bm}
\usepackage{epsfig}
\usepackage{amsfonts}
\usepackage{amssymb,amscd}

\def\lsim{\raise0.3ex\hbox{$<$\kern-0.75em\raise-1.1ex\hbox{$\sim$}}}
\def\gsim{\raise0.3ex\hbox{$>$\kern-0.75em\raise-1.1ex\hbox{$\sim$}}}

\def\beq{\begin{equation}}
\def\eeq{\end{equation}}
\def\bea{\begin{eqnarray}}
\def\eea{\end{eqnarray}}
\def\bq{\begin{quote}}
\def\eq{\end{quote}}

\newcommand{\rr}{\mbox{\boldmath $r$}}

\def\gappeq{\mathrel{\rlap {\raise.5ex\hbox{$>$}}
{\lower.5ex\hbox{$\sim$}}}}

\def\lappeq{\mathrel{\rlap{\raise.5ex\hbox{$<$}}
{\lower.5ex\hbox{$\sim$}}}}

\def\Toprel#1\over#2{\mathrel{\mathop{#2}\limits^{#1}}}

\begin{document}


\title{Inclusive heavy quark photoproduction in $pp$, $pPb$ and $PbPb$ collisions at Run 2 LHC energies }

\author{V.~P. Gon\c{c}alves, G. Sampaio dos Santos, C. R. Sena }
\affiliation{High and Medium Energy Group, \\
Instituto de F\'{\i}sica e Matem\'atica, Universidade Federal de Pelotas\\
Caixa Postal 354, CEP 96010-900, Pelotas, RS, Brazil }


\date{\today}

\begin{abstract}
In this paper we present a comprehensive analysis of  the inclusive heavy quark photoproduction in $pp$, $pPb$ and $PbPb$ collisions at Run 2 LHC energies using the Color Dipole formalism. The rapidity distributions and total cross sections for the charm and bottom production  are estimated considering the more recent phenomenological models for the dipole - proton scattering amplitude, which are based on the Color Glass Condensate formalism and are able to describe the inclusive and exclusive $ep$ HERA data. Moreover, we present, by the first time, the  predictions for the transverse momentum distributions of the $D$ and $B$ mesons produced in the photon -- induced interactions. 

\end{abstract}
\keywords{Ultraperipheral Heavy Ion Collisions, Heavy Quark Production, Meson Production, QCD dynamics}
\pacs{12.38.-t; 13.60.Le; 13.60.Hb}

\maketitle


The advent of the high energy colliders  has allowed us to study the hadron structure at high energies and to achieve a deeper knowledge of the hadronic structure. The investigation of the hadronic structure can be more easily performed in photon -- induced interactions, as those present in the deep inelastic scattering (DIS) process \cite{paul} and in ultraperipheral hadronic collisions \cite{upc}. In particular, photon -- induced interactions can be used to improve our understanding of the strong interactions in the high energy regime \cite{review_forward}. Currently, we knows that gluon density inside the proton grows with the energy and that in this regime the  hadron becomes a dense system and non - linear effects inherent to  the QCD dynamics may become visible. Although our knowledge about the QCD dynamics at high energies have had a substantial development in the last years \cite{hdqcd}, several open questions still remain, which implies that the underlying assumptions of the different approaches should still be tested by the comparison of its predictions with the future experimental data for high energy processes \cite{review_forward,eics}.

During the last years, the LHC has provided data on photon -- induced interactions at  
Run 1 energies \cite{alice,alice2,lhcb,lhcb2,lhcb3,alice3} 
and in this year at Run 2 energies \cite{lhcbconf,aliceconf}.  
One of the more studied processes is the exclusive vector meson photoproduction in 
$pp/pPb/PbPb$ collisions
\cite{klein_prc,gluon,strikman,outros_klein,vicmag_mesons1,
outros_vicmag_mesons,outros_frankfurt,Schafer,vicmag_update,gluon2,motyka_watt,
Lappi,griep,Guzey,Martin,glauber1,bruno1,Xie,bruno3,vicnavdiego,tuchin}, with the basic motivation been associated to the fact that its cross section is proportional to the square of the gluon distribution (in the collinear formalism) \cite{gluon}, being thus strongly dependent on the description of the QCD dynamics at high energies. However, as demonstrated by recent studies \cite{bruno3,nosrecent}, the theoretical uncertainty present in the distinct predictions still is large, which implies that the analysis of a specific final state  probably will not allow us to obtain a final conclusion about the more adequate description of the kinematical range probed by the LHC. Probably, only  the analysis of a wide set of different final states and its description in terms of a unified approach will allows to constrain the main aspects of the QCD dynamics. As a consequence, the analysis of other final states is important and timely. Some possibilities have been discussed e.g. in Ref. \cite{bruno_double} considering exclusive processes, where both incident hadrons remain intact and two rapidity gaps are present in the final state. However, the QCD dynamics also be probed in photon - induced interactions where one the incident hadrons fragments and only one rapidity gap is present in the final state, usually denoted inclusive processes. Examples of inclusive  processes are the heavy quark and dijet photoproduction in hadronic collisions \cite{vogt,vicmaghq,vicmagane,victop,vogtjet,frank,vicmur,kotko,vicmur2}. The main disadvantages of the inclusive processes are: (a) the cross sections are proportional to the first power of the gluon distribution (in the collinear formalism), and (b) the experimental separation becomes harder in comparison to the exclusive one. However, its cross sections are in general one order of magnitude larger. Moreover, recent results obtained by the ATLAS Collaboration \cite{atlas},  indicate that its experimental separation is, in principle, feasible.   Such aspects motivate the analysis of the heavy quark photoproduction in $pp$, $pPb$ and $PbPb$ collisions at the Run 2 LHC energies. In what follows we will improve the studies performed in Refs. \cite{vicmaghq,vicmagane} using the Color Dipole formalism, considering the updated models for the description of the dipole - hadron interaction as well as presenting, for the first time, the predictions for the transverse momentum distributions of the heavy quark and heavy mesons produced in the photon -- induced interactions. As we will show below, the event rates are large, which implies that the analysis of this final state can, in principle, be performed in the Run 2 of the LHC.

\begin{figure}[!t]
\begin{tabular}{cc}
\centerline{{\includegraphics[height=6cm]{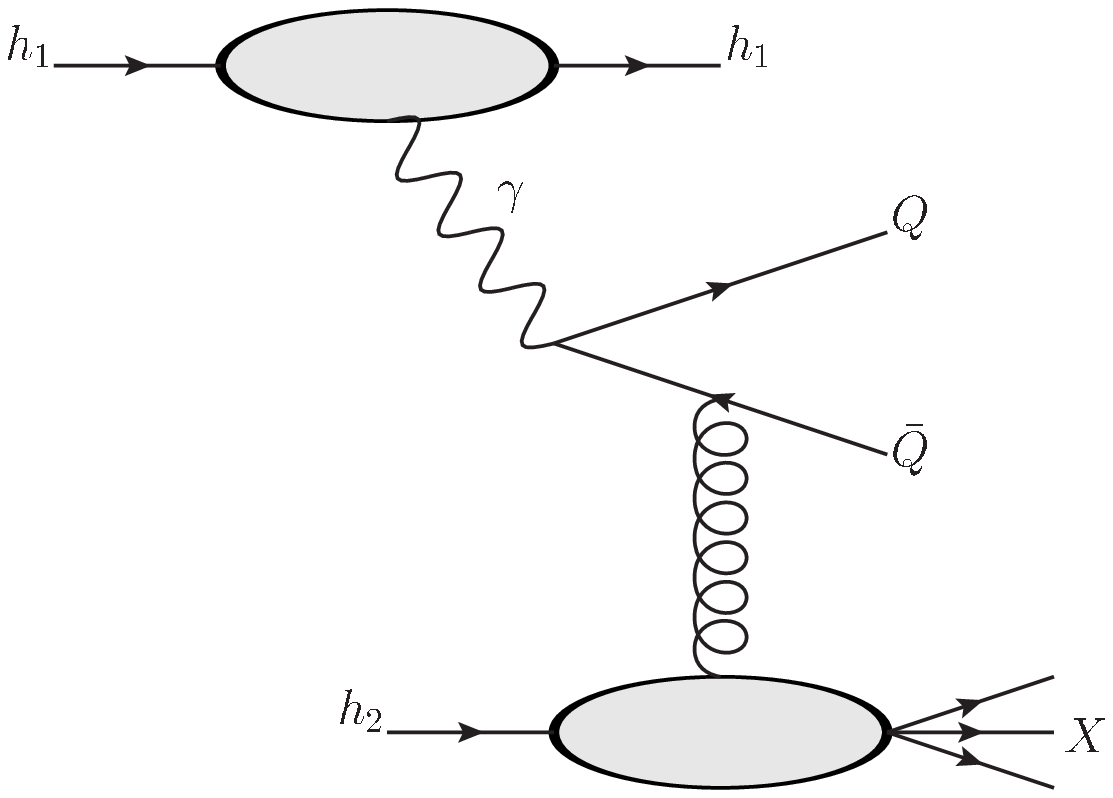}}
{\includegraphics[height=6cm]{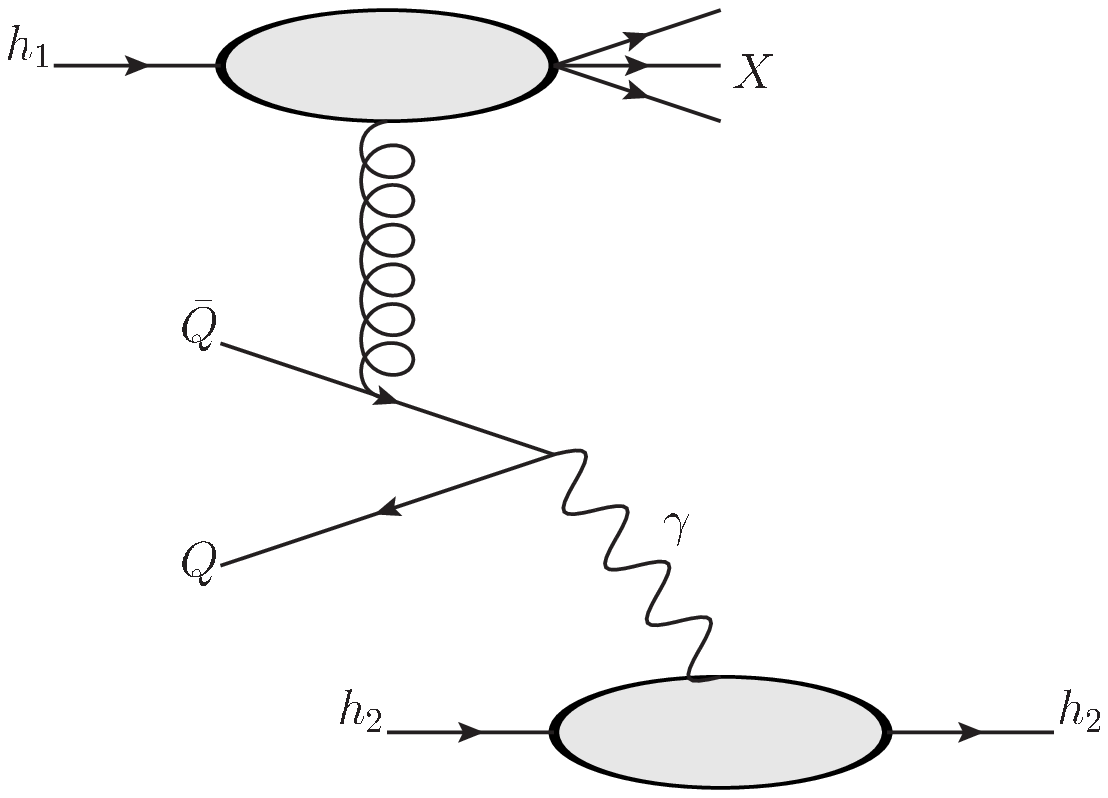}}}
\end{tabular}
\caption{Typical diagrams for the inclusive heavy quark photoproduction in a hadronic collision.}
\label{Fig:diagrama}
\end{figure}

Lets start our study presenting a brief review of the color dipole formalism for the photoproduction of heavy quarks in a ultraperipheral hadronic collision, which is defined by a collision between two hadrons  at impact 
parameters such that $b > R_{1} + R_{2}$, where $R_{i}$ is the radius of the hadron 
$i$. The typical diagrams associated to this process are presented in Fig. \ref{Fig:diagrama}. At high energies, the hadrons act as a source of 
almost real photons and the hadron-hadron cross section can be written 
in a factorized form, described using the equivalent photon approximation (Ref. \cite{upc}). In particular, the differential cross section for the production of a heavy quark $Q\bar{Q}$  at rapidity $Y$ with the quark having transverse momentum $p_T$ will be given by
\begin{eqnarray}
\frac{d^2\sigma \,\left[h_1 + h_2 \rightarrow   h_1 + Q\bar{Q} + h_2\right]}{dYd^2p_T} = \left[n_{h_{1}} (\omega) \,\frac{d\sigma_{\gamma h_2 \rightarrow Q\bar{Q} h_2}}{d^2p_T}\left(W_{\gamma h_2}^2 \right)\right]_{\omega_L} + \left[n_{h_{2}} (\omega)\,\frac{d\sigma_{\gamma h_1 \rightarrow  Q\bar{Q} h_1}}{d^2p_T}\left(W_{\gamma h_1}^2 \right)\right]_{\omega_R}\,\,,
\label{dsigdy}
\end{eqnarray}
where $\omega_L \, (\propto e^{+Y})$ and $\omega_R \, (\propto e^{-Y})$ denote photons from the $h_1$ and $h_2$ hadrons, respectively. Moreover, $n(\omega)$ is the equivalent photon spectrum generated by  the 
hadronic source and $d\sigma/d^2p_T$ is the differential cross section for the production of a heavy quark $Q$ with transverse momentum $p_T$ in a photon - hadron interaction with center - of - mass energy $W_{\gamma h} = \sqrt{4 \omega E}$, where $E = \sqrt{s}/2$ and $\sqrt{s}$ is the hadron-hadron c.m. energy. The final state will be characterized by one rapidity gap, associated to the photon exchange, and an intact hadron in the final state, which was the photon source. As in our previous studies \cite{vicnavdiego,bruno3}, we will assume that the  photon flux associated to the proton and nucleus can be described by the Drees - Zeppenfeld \cite{Dress} and the relativistic point – like charge \cite{upc} models, respectively.

The heavy quark photoproduction cross section will be estimated using the Color Dipole formalism \cite{nik}, which allows us to study  the $\gamma h$ interaction in terms of a  (color) dipole - hadron interaction and  take into account of the non - linear effects in the QCD dynamics. In this formalism, the photon - hadron cross sections are given in terms of the photon wave function $\Psi$, which describes the photon fluctuation into a color $Q \bar{Q}$ dipole which interacts with the target via strong interaction, with this interaction being described by the dipole - hadron cross section $\sigma_{dh}$. In particular, the transverse momentum distribution will be given by \cite{kopel,wolfnik}
\begin{eqnarray}  
\frac{d\sigma(\gamma h \rightarrow Q\bar{Q} X)}{d^{2}p_{T}} &=& \frac{1}{(2\,\pi)^{2}} 
\int d^{2}\mathbf{r_{1}}\,d^{2}\mathbf{r_{2}}\,d\alpha\,\textrm{e}^{i\mathbf{p_{T}}\cdot(\mathbf{r_{1}}-\mathbf{r_{2}})}\,  
\Psi^{T}(\alpha,\mathbf{r_{1}})\Psi^{* T}(\alpha,\mathbf{r_{2}})\nonumber \\  
&\times& \frac{1}{2}\left\{\sigma_{dh}(x,\mathbf{r_{1}})+  
\sigma_{dh}(x,\mathbf{r_{2}})-\sigma_{dh}(x,\mathbf{r_{1}}-\mathbf{r_{2}})\right\}\,\,,
\label{eq8}  
\end{eqnarray}
where  $\alpha$ is the photon momentum fraction carried by the quark and  $\mathbf{r_{1}}$ and $\mathbf{r_{2}}$ are the transverse dipole separations in the amplitude and its complex conjugate, respectively. As shown in Refs. \cite{kopel,wolfnik}, for a transversely polarized photon with $Q^2 = 0$ one have that the overlap function $\Psi^{T}(\alpha,\mathbf{r_{1}})\Psi^{* T}(\alpha,\mathbf{r_{2}})$ is given by
\begin{eqnarray}   
\Psi^{T}(\alpha,\mathbf{r_{1}})\Psi^{* T}(\alpha,\mathbf{r_{2}}) &=& \frac{6\,\alpha_{em}\,e_{Q}^{2}}{(2\,\pi)^{2}}\left\{m_{Q}^{2}K_{0}  
(m_Q\, r_{1})K_{0}(m_Q\, r_{2})  
+ m_Q^{2}[\alpha^{2}+(1-\alpha)^{2}]  
\frac{\mathbf{r_{1}}\cdot \mathbf{r_{2}}}{r_{1}r_{2}}K_{1}(m_Q\, r_{1})
K_{1}(m_Q\, r_{2})\right\} \,,
\label{eq9}
\end{eqnarray}  
where $e_Q$ is the fractional quark charge and $m_Q$ the mass of the heavy quark.
Furthermore, $x = 4m_Q^2 / W_{\gamma h}^2$ and the dipole -- hadron cross section can be expressed by
\begin{eqnarray}
 \sigma_{dh}(x,r^{2}) = 2 \int d^{2} \textbf{\textit{b}}_{h} \,\, {\cal N}_{h} 
(x,\textbf{\textit{r}},\textbf{\textit{b}}_{h}) ,
\end{eqnarray}
where $\textbf{\textit{b}}_{h}$ is the impact parameter, given by the transverse distance  between the dipole center  and the target center, and ${\cal N}_{h} (x,\textbf{\textit{r}},\textbf{\textit{b}}_{h})$ is the forward dipole - hadron scattering amplitude, which is dependent  on the modelling of the QCD dynamics at high energies (See below). As demonstrated in Ref. \cite{kopel},  the substitution of Eq. (\ref{eq9}) into (\ref{eq8}) allow us  to express the spectra in terms of integrals over the longitudinal momentum $\alpha$ and the dipole size $r$ as follows
\begin{eqnarray}  
\frac{d\sigma(\gamma h \rightarrow Q\bar{Q} X)}{d^{2}p_{T}}  
&=&\frac{6\,e_{Q}^{2}\,\alpha_{em}}{(2\,\pi)^{2}}\int d\alpha
\left\{m_{Q}^{2}  \left[\frac{I_{1}}{p_{T}^{2}+m_Q^{2}}-\frac{I_{2}}{4\,m_Q}\right]  + 
\left[\alpha^{2}+(1-\alpha)^{2}\right] \left[\frac{p_{T}\,m_Q\, I_{3}}{p_{T}^{2}+m_Q^{2}}
-\frac{I_{1}}{2}+\frac{m_Q\, I_{2}}{4}\right]\right\} \,\,,  
\label{eq10}  
\end{eqnarray}  
where we have defined the auxiliary functions   
\begin{eqnarray}  
I_{1}&=&\int dr\,r\,J_{0}(p_{T}\,r)\,K_{0}(m_Q\, r)\,\sigma_{dh}(r) \label{eq11} \\  
I_{2}&=&\int dr\,r^{2}\,J_{0}(p_{T}\,r)\,K_{1}(m_Q\, r)\,\sigma_{dh}(r) \label{eq12} \\  
I_{3}&=&\int dr\,r\,J_{1}(p_{T}r)\,K_{1}(m_Q\, r)\,\sigma_{dh}(r)\,\,,  
\label{eq13}  
\end{eqnarray}  
with the functions $K_{0,1}$ ($J_{0,1}$) being the modified Bessel functions of the second (first) kind.

In the Color Dipole formalism the main ingredient for the calculation of the heavy quark cross sections is the dipole - target scattering amplitude ${\cal N}_h$. The treatment of this quantity for the nucleon and nuclear case is the subject of intense study by several groups \cite{hdqcd}.  In the case of a proton target, the color dipole 
formalism has been extensively used to describe the inclusive and exclusive HERA data. During the last decades, several phenomenological models based on the Color Glass Condensate formalism \cite{hdqcd} have been proposed to describe the HERA data taking into account the non - linear effects in the QCD dynamics. In general, such models differ in the treatment of the impact parameter dependence and/or of the linear and non - linear regimes. Two examples of very successful models are the   b-CGC \cite{Watt_bCGC,KMW} and IP-SAT \cite{ipsat1} models, which  have been updated in Refs. \cite{Rezaeian_update,ipsat4}, using the high precision HERA data to constrain their 
free parameters, and describe the data quite well. The b-CGC model interpolates two analytical 
solutions of well known evolution equations: the solution of the BFKL equation near  the 
saturation regime and the solution of the  Balitski-Kovchegov 
equation deeply inside the saturation regime. Moreover, it assumes that the saturation scale depends on the impact parameter. On the other hand, 
  IP-SAT model \cite{ipsat1,ipsat2,ipsat3} assumes an eikonalized 
form for ${\cal N}_p$  that depends on a gluon distribution evolved via DGLAP equation and the proton profile in the impact parameter space. Although both the b-CGC and the IP-SAT models include saturation effects in the description of the QCD dynamics and depend on impact-parameter, the underlying dynamics of two models is quite different. While the b-CGC model probes saturation through the increasing of the gluon density driven by the BFKL evolution, in the case of the IP-SAT model such increasing is driven by the DGLAP one. In what follows we will use in our calculations the updated versions of these models presented in Refs. \cite{Rezaeian_update,ipsat4}. In the nuclear case, we will assume the model proposed in Ref. 
\cite{armesto}, which is based on the Glauber - Mueller approach \cite{mueller}, includes the  impact parameter dependence and describes  
the existing  experimental data on the nuclear structure function \cite{erike}. 
In this model the  dipole-nucleus scattering amplitude is given by
\begin{eqnarray}
 {\cal N}_{A}(x,\textbf{\textit{r}},\textbf{\textit{b}}_{A}) = 1 - \exp \left[
-\frac{1}{2} \sigma_{dp}(x,r^{2}) \, T_{A}(\textbf{\textit{b}}_{A})
 \right] ,
 \label{Na_Glauber}
\end{eqnarray}
where $T_{A}(\textbf{\textit{b}}_{A})$ is the nuclear thickness, 
 which is obtained from a 3-parameter Fermi distribution for the nuclear
density normalized to $A$.
The above equation
sums up all the 
multiple elastic rescattering diagrams of the $Q \overline{Q}$ pair
and is justified for large coherence length, where the transverse separation $\rr$ of 
partons in the multiparton Fock state of the photon becomes a conserved quantity, {\it i.e.} the size of the pair $\rr$ becomes eigenvalue
of the scattering matrix.  In what follows we will compute  $\mathcal{N}_A$ considering 
the b-CGC and IP-SAT models for the dipole - proton scattering amplitude discussed before.


\begin{figure}[!t]
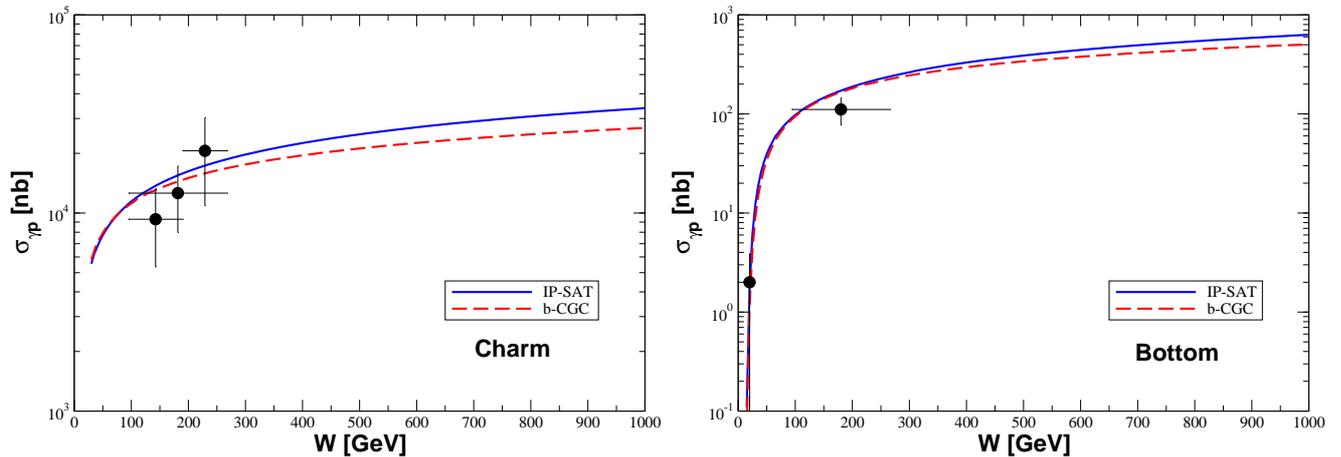

\begin{tabular}{cc}
\centerline{{\includegraphics[height=6cm]{sigtot_charm_q2_0_mc_1_27.eps}}
{\includegraphics[height=6cm]{sigtot_bottom_q2_0_mb_4_5.eps}}}
\end{tabular}
\caption{Comparison between the b-CGC and IP-SAT predictions and the HERA data \cite{heradata} for the charm (left panel) and bottom (right panel) photoproduction.}
\label{Fig:HERA}
\end{figure}

Initially, lets compare the b-CGC and IP-SAT predictions with the  HERA data \cite{heradata} for the  total charm and bottom photoproduction.  The results, calculated assuming $m_c = 1.27$ GeV and $m_b = 4.5$ GeV, are presented in Fig. \ref{Fig:HERA}. One have that in the HERA kinematical range, the b-CGC and IP-SAT predictions are similar. On the other hand, at higher energies, the IP-SAT predictions are larger than the b-CGC one, which is directly associated to different treatment of the saturation regime present in these models. The difference is larger in the charm case, since for this final state the contribution of the saturation effects is larger than in the bottom case. In what follows we will analyze the impact of these differences in the modelling of the QCD dynamics at high energies on the heavy quark production in ultraperipheral collisions at the LHC.

\begin{figure}[t]
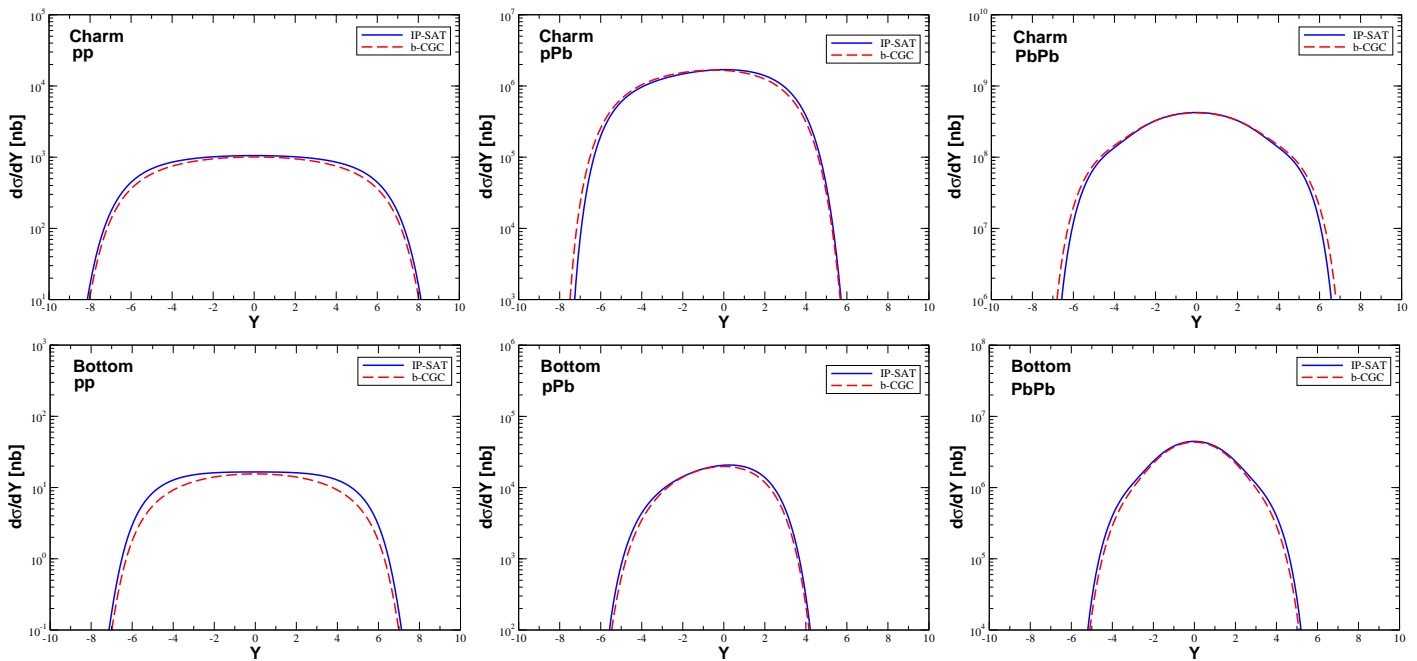

  \begin{tabular}{ccc}
    {\psfig{figure=dsigdy_charm_pp_13000.eps,scale=0.25}} 
    & {\psfig{figure=dsigdy_charm_pPb_8100.eps,scale=0.25}} & {\psfig{figure=dsigdy_charm_PbPb_5020.eps,scale=0.25}} \\
    {\psfig{figure=dsigdy_bottom_pp_13000.eps,scale=0.25}} 
    & {\psfig{figure=dsigdy_bottom_pPb_8100.eps,scale=0.25}} & {\psfig{figure=dsigdy_bottom_PbPb_5020.eps,scale=0.25}} 
  \end{tabular}
  \caption{Rapidity distributions for the charm (upper panels) and bottom (lower panels) photoproduction in $pp$ ($\sqrt{s} = 13$ TeV), $pPb$ ($\sqrt{s} = 8.1$ TeV) and $PbPb$ ($\sqrt{s} = 5.02$ TeV) collisions at the LHC. }
  \label{Fig:rapidity}
\end{figure}

In Fig. \ref{Fig:rapidity} we present our predictions for the rapidity distributions for the inclusive charm and bottom photoproduction in $pp$, $pPb$ and $PbPb$ collisions at the Run 2 LHC energies. We obtain that the nuclear distributions are enhanced by a $Z^2$ factor, present in the nuclear photon flux. Moreover, the predictions for $pPb$ collisions are asymmetric in rapidity due to the asymmetry on the initial photon fluxes associated to a proton and a nucleus, with the $\gamma h$ interactions being dominated by photons generated by the nucleus.  It is important to emphasize that the overlap function for the charm  is dominated by  larger dipole sizes than for the bottom case. Therefore, the charm and bottom quark production  probe ${\cal N}_h$ at different values of $\rr$. We obtain that the b-CGC and IP-SAT predictions are similar, with small differences at larger rapidities, where larger values of photon - hadron center - of - mass energies are probed. The results indicate that the uncertainty present in the color dipole predictions for the heavy quark production is small, which implies that a future measurement of this observable is an important probe of this approach.  The corresponding values for the total cross sections are presented in Table 
\ref{Tab:sectot} considering two rapidity ranges. In particular, we show our results for the rapidity range analyzed by the LHCb experiment, which probes the heavy quark photoproduction at forward rapidities ($2 < Y < 4.5$), where we expect a larger contribution of the non - linear effects.  One have that the predictions for the LHCb range are approximately one order of magnitude smaller than if the full kinematical range is considered. In the case of $pp$ and $pPb$ collisions, the difference between the b-CGC and IP-SAT is $\approx 10\% \, (20 \%)$ for the charm (bottom) production. For $PbPb$ collisions, this difference is smaller, which is directly associated to the fact that we are using the Eq. (\ref{Na_Glauber}) to describe the dipole - nucleus scattering, with the b-CGC and IP-SAT only affecting the argument of the exponential.

\begin{table}[t]
\centering
\begin{tabular}{|c|c|c|c|}\hline \hline
      &   {\bf Rapidity range} &                {\bf b-CGC}                         &                 {\bf IP-SAT}                               \\ \hline \hline        
{\bf pp ($\sqrt{s} = 13$ TeV)} & $-10 < Y < 10$  &  1.03$\times10^{4}$ nb (1.31$\times10^{2}$ nb)  &  1.14$\times10^{4}$ nb (1.60$\times10^{2}$ nb)       \\ \hline
  &  $2 < Y < 4.5$ & 2.08$\times 10^{3}$ nb (2.81$\times 10^{1}$ nb)  &  2.30$\times 10^{3}$ nb (3.56$\times 10^{1}$ nb)       \\ 
  \hline
\hline
{\bf pPb ($\sqrt{s} = 8.1$ TeV)} &  $-10 < Y < 10$  & 1.21$\times10^{7}$ nb (9.84$\times10^{4}$ nb)  &  1.22$\times10^{7}$ nb (1.07$\times10^{4}$ nb)       \\
 \hline
  &  $2 < Y < 4.5$ &  1.70$\times 10^{6}$ nb (9.39$\times 10^{3}$ nb)  &  1.98$\times 10^{6}$ nb (11.50$\times 10^{3}$ nb)       \\ \hline
  \hline
{\bf PbPb ($\sqrt{s} = 5.02$ TeV)} & $-10 < Y < 10$  & 2.81$\times10^{9}$ nb (1.87$\times10^{7}$ nb)  &  2.75$\times10^{9}$ nb (1.99$\times10^7$ nb)          \\ \hline
  &  $2 < Y < 4.5$ &  5.29$\times 10^{8}$ nb (2.23$\times 10^{6}$ nb)  &  5.12$\times 10^{8}$ nb (2.56$\times 10^6$ nb) \\
  \hline
  \hline  
\end{tabular} 
\caption{Total cross sections for the inclusive charm (bottom) photoproduction in $pp/pPb/PbPb$ collisions at the Run 2 LHC energies considering two rapidity ranges.}
\label{Tab:sectot}
\end{table}

We predict large values for the event rates, in particular for the charm production in $PbPb$ collisions. One important aspect, that deserves a more detailed study, is the impact of this large number of charm quarks produced in photon -  induced interactions in the description of the $D$ meson production in nuclear collisions, especially in peripheral collisions  ($b \approx 2 R_{Pb}$). As recently observed by the ALICE Collaboration \cite{alice_per}, in order to describe the $J/\Psi$ production at small transverse momentum in peripheral collisions we should to take into account the contribution of the $J/\Psi$ production in photon -- nucleus interactions (For a first theoretical discussion about the subject see Ref. \cite{antonimariola}).   Considering our predictions for the heavy quark production, we can expect that a similar effect should also be present in the $D$ and $B$ meson production in peripheral $PbPb$ collisions. Surely, this theme should be investigated in more detail in the future.   

Lets now estimate, for the first time, the transverse momentum distributions of the heavy quarks (and mesons) produced in UPHIC at the LHC. The distribution for heavy quarks can be directly obtained in the color dipole formalism using Eqs. (\ref{dsigdy}) and (\ref{eq8}). On the other hand, in order to calculate the momentum spectra for heavy mesons we need to take into account the hadronization of the heavy quarks through the corresponding fragmentation function, which is associated to the probability of a heavy quark to generate a given heavy meson. We will focus our analysis on the production of $D^0$ and $B^0$ mesons, but it can be easily extended for other final states. We have that
\begin{eqnarray} 
\frac{d^{2}\sigma(h_1+h_2 \rightarrow H + X)}{dY_Hdp^{2}_{T,H}} = \int_{z_{\mathrm{min}}}^1 \frac{dz}{z^2} \, 
\mathcal{D}^{Q/H}(z,\mu^2) \, \left[\frac{d^{2}\sigma(h_1+h_2 \rightarrow Q\bar{Q}+X)}{dY_Qdp^{2}_{T,Q}}\right]_{p_{T,Q} = \frac{p_{T,H}}{z}} \,\,,
\label{eq15}
\end{eqnarray}
where $p_{T,H}$ is the transverse momentum of the heavy meson, $z$ is the fractional light-cone momentum of the heavy quark $Q$ carried by
the meson $H$ and $\mathcal{D}^{Q/H}(z,\mu^2)$ is the fragmentation function at the scale $\mu^2$. Moreover, we  have  made the  typical  approximation  assuming  that the heavy quark rapidity is  unchanged  in  the fragmentation process, i.e. $Y_H = Y_Q$ \cite{rafal}. In our  calculations  we  use  standard  Peterson  model
of  fragmentation function \cite{peterson}, which is given by
\begin{eqnarray}  
\mathcal{D}^{Q/H}(z,\mu^2) = \frac{n(H)}{z\left[1-\frac{1}{z}-\frac{\epsilon_{Q}}{1-z}\right]^{2}} \,\,,
\label{eq16}  
\end{eqnarray}  
 with $n(H)$ being obtained by the normalization of the fragmentation functions to the branching fractions and the parameter $\epsilon_Q$ is assumed to be $\epsilon_{c} = 0.05$ and $\epsilon_{b} = 0.006$. During the last years, several authors have proposed  harder fragmentation functions  (with smaller values of $\epsilon_Q$) as well as have studied the effects of the QCD evolution \cite{metz}. However, the results presented e.g. in Refs. \cite{rafal,vicboris} indicate that in the range of small values of the transverse momentum, these different models predict similar distributions. Considering that this is the range of interest in our analysis, we will perform our calculations, for simplicity, using the Peterson model.

\begin{figure}[t]
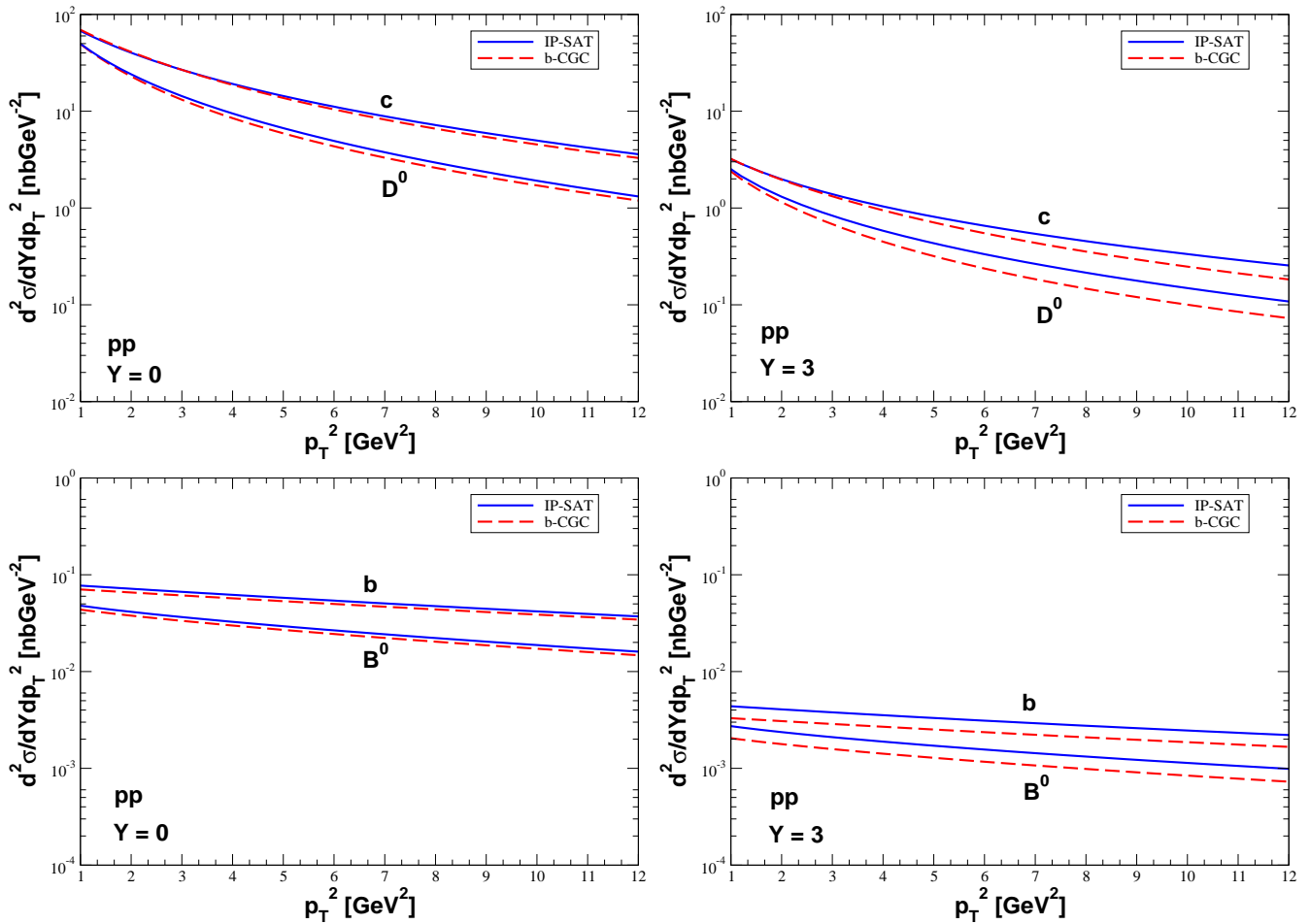

  \begin{tabular}{cc}
    {\psfig{figure=dsigdpt_charm_D0_pp_13000.eps,scale=0.35}} 
 & {\psfig{figure=dsigdpt_charm_D0_pp_13000_y_3.eps,scale=0.35}} \\
    {\psfig{figure=dsigdpt_bottom_B0_pp_13000.eps,scale=0.35}}  & {\psfig{figure=dsigdpt_bottom_B0_pp_13000_y_3.eps,scale=0.35}} 
  \end{tabular}
  \caption{Transverse momentum distributions for the photoproduction of heavy quarks and heavy mesons in $pp$ collisions at $\sqrt{s} = 13$ TeV considering two typical values for the rapidity.}
  \label{Fig:spectrapp}
\end{figure}

\begin{figure}[t]
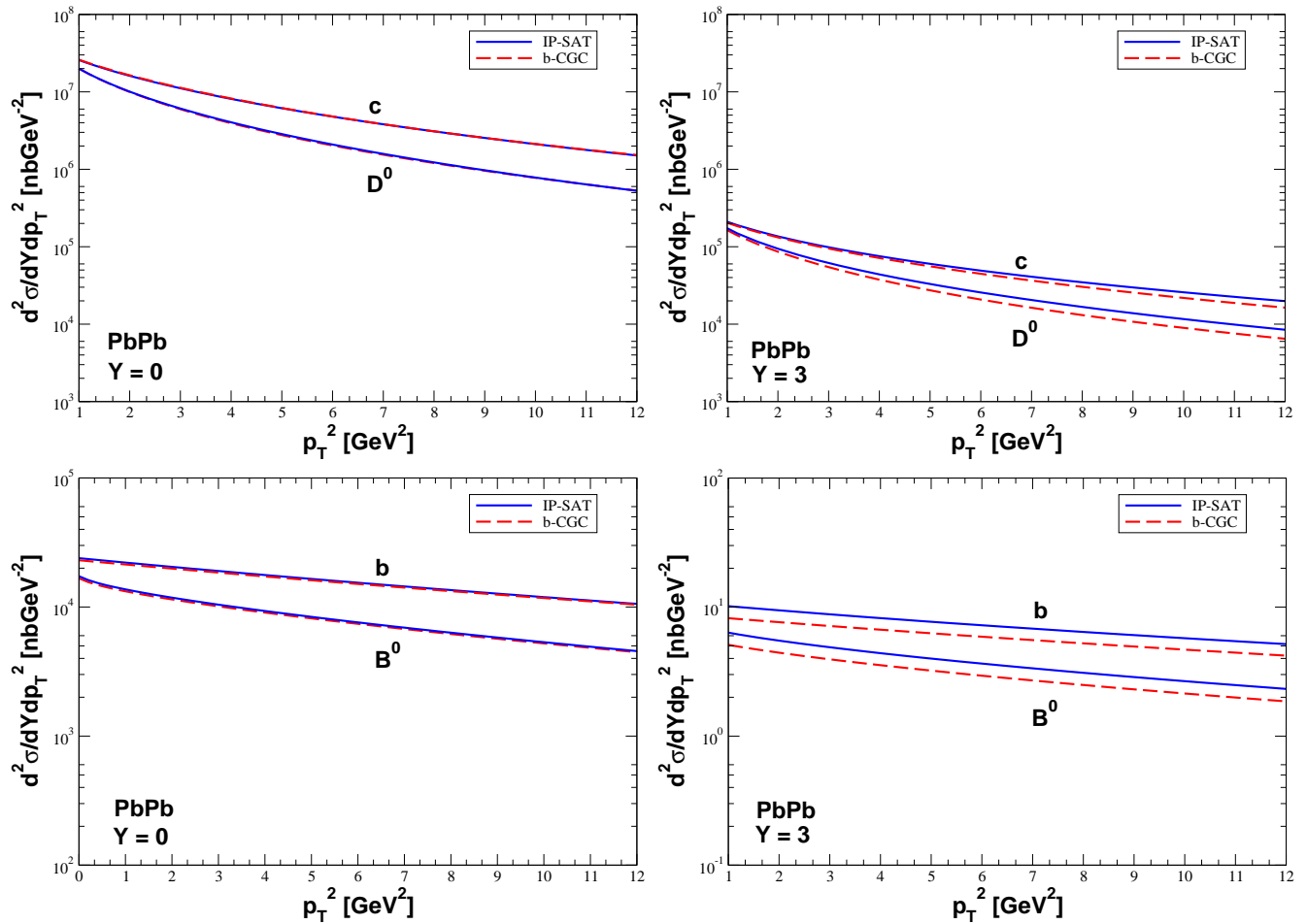

  \begin{tabular}{cc}
    {\psfig{figure=dsigdpt_charm_D0_PbPb_5020.eps,scale=0.35}} 
 & {\psfig{figure=dsigdpt_charm_D0_PbPb_5020_y_3.eps,scale=0.35}} \\
    {\psfig{figure=dsigdpt_bottom_B0_PbPb_5020.eps,scale=0.35}}  & {\psfig{figure=dsigdpt_bottom_B0_PbPb_5020_y_3.eps,scale=0.35}} 
  \end{tabular}
  \caption{Transverse momentum distributions for the photoproduction of heavy quarks and heavy mesons in $PbPb$ collisions at $\sqrt{s} = 5.02$ TeV considering two typical values for the rapidity.}
  \label{Fig:spectraPbPb}
\end{figure}

Our predictions for the transverse momentum spectra of the charm and bottom quarks and $D^0$ and $B^0$ mesons produced in photon -- proton and photon - nucleus interactions in $pp$ and $PbPb$ collisions at the LHC are presented in Figs. \ref{Fig:spectrapp} and \ref{Fig:spectraPbPb}, respectively, considering two different values for the rapidity. The distributions for $pPb$ collisions differ only in magnitude  to those obtained for $pp$ collisions, being enhanced by a factor $Z^2$, but are similar in its $p_T$ behaviour, since both are generated by $\gamma p$ interactions. It is important to emphasize that we use the common label $p_T^2$ in the horizontal axis, but it refers to the heavy quark (meson)  transverse momentum when we are discussing the charm and bottom ($D^0$ and $B^0$) results, and $p_{T,Q} \neq p_{T,H}$ as indicated in Eq. (\ref{eq15}). Initially, lets discuss our results for $pp$ collisions presented in Fig. \ref{Fig:spectrapp}. We have that, for a fixed $p_T$, the  distributions decrease when the rapidity is increased, which is expected from the results shown in Fig. \ref{Fig:rapidity}. Moreover, the inclusion of the fragmentation modifies the behaviour of the heavy quark distributions, with the corresponding meson distributions having a smaller magnitude and decreasing faster with $p_T$. The bottom/$B^0$ distributions are flatter than the charm/$D^0$ one due to the larger quark mass. Additionally, at central rapidities ($Y = 0$), the b-CGC and IP-SAT predictions are very similar. On the other hand, they start to be different when the rapidity is increased to $Y = 3$, which is directly associated to the distinct treatment of the QCD dynamics present in these models. The results for $PbPb$ collisions presented in Fig. \ref{Fig:spectraPbPb} are similar to the $pp$ case, with the difference between the b-CGC and IP-SAT predictions being still smaller, as expected from our analysis of the rapidity distributions.
Our results indicate that the uncertainty present on the color dipole predictions for the heavy quark photoproduction in hadronic collisions is small, which implies that a future experimental analysis of this final state can be useful to probe this formalism and its underlying assumptions.

Finally, lets summarize our main results and conclusions. The recent results for photon - induced interactions in hadronic colliders has indicated that the analysis of these processes can be useful to improve our understanding of the strong interactions, in particular about the treatment of the QCD dynamics at high energies. The forthcoming Run 2 of the LHC will provide a larger data sample, allowing the  study of a  
larger set of different final states and  a better discrimination between 
alternative descriptions. This possibility has motivated the analysis performed in this paper, where  we have presented a comprehensive study of the inclusive heavy quark and heavy meson photoproduction in $pp/pPb/PbPb$ collisions at Run 2 LHC energies 
using the  color dipole formalism.  We have used the updated versions of different 
models of the dipole scattering amplitude, which take into account   the 
non - linear effects of the QCD dynamics (which are expected 
to become visible at the currently available energies) and describe the  HERA data for inclusive and exclusive processes. As the LHC probes a larger range of $\gamma h$ center of mass energies, the analysis of   the inclusive heavy quark photoproduction in this collider can be useful to probe the color dipole formalism and its underlying assumptions. As the free parameters present in the color dipole formalism have been    
constrained by the HERA data, the predictions for LHC energies are parameter  
free. In our study we have presented predictions for the photoproduction of charm, bottom, $D^0$ and $B^0$ in $pp/pPb/PbPb$ collisions. We predict large values for the event rates at the LHC. The predictions for the transverse momentum distributions have been presented by the first time. Our results  
demonstrated that the uncertainty present in the color dipole predictions is small, which implies that the analysis of this process can test the universality of the color dipole description for inclusive and exclusive processes.
Future experimental data may decide whether improvements of the color  
dipole description should also be included in the analysis of the photon -- induced interactions.


\section*{Acknowledgements}
 This work was partially financed by the Brazilian funding agencies CAPES, CNPq,  FAPERGS and INCT-FNA (process number 464898/2014-5).



\end{document}